\documentclass[a4paper,twoside]{article}
\usepackage{AMS2LA,fancyhdr,CJK,xcolor,multicol,graphics,greekup,bm,cmap,epstopdf}
\usepackage{amssymb}\usepackage{gensymb,mathcomp}
\usepackage[pdfstartview=FitH, bookmarks=false, colorlinks=false,
    linkcolor=blue, urlcolor=blue, pdfborder=001, citecolor=blue,
    pdftitle={Changepdftitle}, pdfauthor={Changepdfauthor},
    pdfcreator=CHIN. PHYS. LETT., pdfproducer=CPL,
    pdfkeywords={changePACS}]{hyperref}
\usepackage{graphicx}
\newcommand{\explel}{{\large\sffamily~~~~~~~~~~~~~~~~~} \hfill}
\newcommand{\expler}{\hfill\fcolorbox{lightgray}{lightgray}{\textcolor{red}{{\large\fontfamily{phv}\textbf{New Submission}}}}}

\def\headrule{\kern 1mm \hrule width 17cm \kern -1mm}%
\def\footnoterule{\kern 1mm \hrule width 7cm \kern 2.2mm}%
\def\REF#1{\par\hangindent\parindent\indent\llap{#1\enspace}\ignorespaces}%
\newcommand{\ucite}[1]{$^{[#1]}$}
\renewcommand{\cite}[1]{\,[#1]}

\newcommand{\cplyear}{2020} \newcommand{\cplvol}{37}
\newcommand{\cplno}{x} \newcommand{\cplpagenumber}{xx{xxxx}}
 \newcommand{\cplpage}{\cplpagenumber-\thepage}

\begin{document}
\begin{CJK}{GBK}{song}\vspace* {-6mm} \begin{center}
%-------------------------Title----------------------------
\large\bf{\boldmath{From finite nuclei to neutron stars: the essential role of high-order density dependence in effective forces}}
%------------------------Footnote--------------------------
%------------------------Footnote--------------------------
\footnotetext{\hspace*{-5.4mm}Supported by the National Key R{\&}D Program of China (Grant No.~2018YFA0404403), the National Natural Science Foundation of China (Grant Nos.~11975032, 11835001, 11790325, 11961141003.).

\noindent$^{*}$Corresponding author. Email: peij@pku.edu.cn

%cplexpress

\noindent\copyright\,{\cplyear}
\href{http://www.cps-net.org.cn}{Chinese Physical Society} and
\href{http://www.iop.org}{IOP Publishing Ltd}}
\\[6mm]
%------------------------Authors----------------------------
\normalsize \rm{} C.J. Jiang, Y. Qiang, D.W. Guan, Q.Z. Chai, C.Y. Qiao, and J.C. Pei$^{*}$
%----------------------Affiliations-------------------
\\[2mm]\small\sl State Key Laboratory of Nuclear Physics and Technology, School of Physics, Peking University, Beijing 100871, China
%------------------------Received date----------------------
\\[4mm]\normalsize\rm{}(Received 28 December 2020; accepted xxx; published online )
\end{center}
\end{CJK}
%----------------------Abstract and PACS--------------------
\vskip 1.5mm

%Email: 1700011347@pku.edu.cn£» peij@pku.edu.cn
%Cell-phone: 18811518133£»18210090960

\noindent{\narrower\small{}
A unified description of finite nuclei and equation of state of neutron stars present a major challenge as well as opportunities for understandings of nuclear interactions.
Inspired by the Lee-Huang-Yang formula of hard-sphere gases, we developed effective nuclear interactions with an additional high-order density dependent term.
The original Skyrme force SLy4 is widely used in studies of neutron stars but is not satisfied for global descriptions of finite nuclei.
The refitted SLy4${'}$ force can improve descriptions of finite nuclei but slightly reduces the radius of neutron star of 1.4 solar mass.
We found that the extended SLy4 force with a higher-order density dependence can properly describe properties of both finite nuclei and GW170817 binary neutron stars, including the mass-radius relation and the tidal deformability. This demonstrated the essential role of high-order density dependence at ultrahigh densities.
 Our work provides a unified and predictive model for  neutron stars, as well as new insights for the future development of effective interactions.

\par}\vskip 3mm
\normalsize\noindent{\narrower{PACS: 21.65.+f; 26.60.+c; 21.30.-x; 21.10.Dr}
%%%CheckPACS: 42.82.-m  Integrated optics
%%%CheckPACS: 42.79.-e  Optical elements, devices, and systems (for integrated optics, see 42.82.-m; for fiber optics, see 42.81.-i) ...    Optical instruments, equipment and techniques, see 07.60.-j and 07.57.-c ...    Optical spectrometers, see 07.57.Ty and 07.60.Rd ...    Photography, photographic instruments and techniques, see 07.68.+m ...   Magnetooptical devices, see 85.70.Sq
{\rm\hspace*{13mm}DOI: 10.1088/0256-307X/\cplvol/\cplno/\cplpagenumber}

\par}\vskip 6mm
%-------------------TEXT TEXT TEXT TEXT---------------------
\begin{multicols}{2}
\emph{Introduction.}--- Neutron stars (NSs), which are compact objects born in the supernova explosion,\ucite{1} attract nuclear physicists' strong interests as they are a unique laboratory for studying the properties of  ultradense
nuclear matter.\ucite{2,3}
The astrophysical observables of NS' mass, radius and the mass-radius relation can provide constraints on nuclear theories extrapolated from properties of finite nuclei.\ucite{4} On 17 August 2017, the gravitational waves emitted from the merging binary NS GW170817\ucite{5} are detected by the Advanced LIGO and the Advanced Virgo,  marking a new era of multi-messenger astronomy. The GW170817 event not only extends our knowledge of NSs' mass and radius, but also provides a new astronomical observable, the quadrupolar tidal deformability\ucite{6}, as a further constraint.\ucite{7-9}

The theoretical descriptions of NS's mass, radius and tidal deformability depend solely on the equation of state (EOS) of nuclear matter.\ucite{10-13}
Therefore, to seek a unified description of properties of finite nuclei and EoS of neutron stars is a major challenge.
Because Quantum Chromodynamics (QCD) is non-perturbative at low energies, it is in principle possible to obtain realistic model-independent nuclear forces from Lattice QCD calculations.\ucite{14}  However, such kind of precise calculations are
so formidable that can be accomplished by today's supercomputers.
The chiral effective field theory based on QCD can derive accurate realistic forces for light nuclei but can propagate large uncertainties towards heavy nuclei and dense nuclear matter, due to the less constrained many-body forces.\ucite{15} To this end, various effective nuclear interactions are extensively explored for EOS of neutron stars.\ucite{16,17}
The effective interactions are developed by taking into account the medium effects, i.e., dependent on surrounding nuclear densities, in contrast to realistic forces in free spaces.
The effective interactions are rather well constrained around the saturation density with experimental studies of finite nuclei.
However, the extrapolation of effective interactions is very risky to neutron stars which could be several times denser than the saturation density.
There are experiments to study the correlation between finite nuclei and neutron stars by measurements of neutron skins of neutron-rich nuclei, combined with
model dependent analysis.\ucite{18,19} It is expected that improved measurements of symmetry energies via heavy ion collisions can directly study EOS at high densities.\ucite{20}

In this Letter, we concern about the applicability of the widely used Skyrme forces for neutron stars.
The Skyrme force is a low-momentum effective forces which is very successful in descriptions of structures
and dynamics of the whole nuclear chart.\ucite{21,22} Among many Skyrme forces,  the SLy4 parameterization \ucite{23,24} is the very best choice for descriptions of neutron
stars and has been widely used as a reference in astrophysical calculations.\ucite{6} The development of SLy4 has taken into account ab initio UV14+UVII results of EoS of pure neutron matter at high densities.\ucite{23}    Many effective forces with a soft EOS have been excluded by
observations of neutron stars around 2 solar mass.\ucite{25}
The inclusion of hyperons in NS core is questionable due to its soft EOS.\ucite{16,17}
For example, the best descriptions of binding energies of finite nuclei
is obtained by the UNEDF0 parameterization with a global RMSE of 1.5 MeV, however, which is failed to properly describe neutron stars.\ucite{26}
On the other hand, SLy4 force has a global RMSE of 4.37 MeV which is too large.\ucite{27}

In Skyrme forces, it is crucial to include  a repulsive density dependent term to describe the nuclear saturation properties and pressure.\ucite{28}
The density dependent term also simulates the effects of 3-body and many-body forces beyond the low-momentum two-body interactions.
It is understood that the induced 3-body and many-body forces are particularly significant in low-momentum (or soft) interactions,
which would be less significant in realistic (or hard) interactions with a high momentum cutoff.\ucite{29}
In this case, the density dependent term in soft Skyrme forces should be carefully treated.
We speculate that it must be insufficient for Skyrme forces with a simple density dependent term to describe
a variety of nuclear systems from dilute nuclear halos to dense neutron stars.
For systems of hard spheres, its exact energy density functional was derived by Lee, Huang, Yang in
a series of publications.\ucite{30,31} The energy density functional form is written as\ucite{32},
 \begin{equation}
 \begin{array}{ll}
 \varepsilon = & \displaystyle \frac{3\hbar^2}{10m}(3\pi^2)^{2/3}\rho^{5/3}+\frac{\hbar^2\pi a}{m}\rho^2 \vspace{5pt}\\
 &\displaystyle +\frac{2\hbar^2a^23^{4/3}\pi^{2/3}}{35m}(11-2\ln2)\rho^{7/3} \vspace{5pt}\\
 & \displaystyle + 0.78\frac{\hbar^2a^33^{5/3}\pi^{10/3}}{2m}\rho^{8/3} \\
 \end{array}
 \label{leeyang}
 \end{equation}
  in which $a$ denotes a finite radius of particles.
The higher order density dependent terms appear naturally as a result of a
finite radius of particles.  The higher-order correction certainty impacts
the equation of state, which has been proved precisely by experiments of cold atomic gases.\ucite{33}
The finite size effects of nucleons is related to repulsive hard cores in realistic forces but have never been explicitly considered by effective nuclear forces.
The aim of this work is to study the influences of high-order density dependencies in extended Skyrme forces on observables of neutron stars.

\emph{Extended Skyrme forces.}---
The standard Skyrme force is a low momentum effective interaction for Skyrme Hartree-Fock calculations  and time-dependent dynamic calculations.
It includes two-body terms with the momentum expansion to the second order and a density dependent term $\rho^{\gamma}(\vec{r}_1-\vec{r}_2)$.
The density dependent term simulates the three-body and many body correlations. However, the power of density dependence $\gamma$ is largely uncertain.
Inspired by the Lee-Huang-Yang formula, we extended the Skyrme force with an additional higher-order density dependent term.
The energy per nucleon $\frac{E}{A}(\rho)$ in nuclear matter in terms of density $\rho$ is written as,\ucite{27}
\begin{equation}
    \left.
    \begin{aligned}
    \frac{E}{A}(\rho)= & \frac{3\hbar^2}{10m}\left(3\pi^2\right)^{2/3}\rho^{2/3}+\frac{1}{4}t_0(1-x_0)\rho \\
    & +\frac{3}{40}[t_1(1-x_1)+3t_2(1+x_2)]\left(3\pi^2\right)^{2/3}\rho^{5/3} \\
    & +\frac{1}{24}\sum\limits_{i}t_{3i}(1-x_{3i})\rho^{\gamma_i+1}
    \end{aligned}
    \right.
    \label{eq}
\end{equation}
Note that Eq.(2) is for the pure neutron matter. The energy density functional of combined protons and neutrons is given in Ref.[23].
In the extended Skyrme forces, there are two density density dependent terms associated with $t_{3i}(i=1,2)$ and $\gamma_i$.
We take $\gamma_1$=1/6 and $\gamma_2$=1/2 based on SLy4 forces.\ucite{27}
This can be seen as a reasonable and perturbative improvement and it is different from the series of expansions of $\rho^{n/3}$ terms as in Ref.[34].
The pressure is given by $ P(\rho)=\rho^2\frac{\partial\frac{E}{A}(\rho)}{\partial\rho}$. The symmetry energy is given by the energy difference between
pure neuron matter and symmetric nuclear matter.

For simplicity, we only refit the momentum-independent parameters $t_0, t_{3i},x_0, x_{3i}$.
 These momentum-independent parameters are
directly correlated in the $s$-wave channel. Other parameters are kept as the same as SLy4 forces.
We obtained different parameter sets by fitting binding energies and charge radii of groups of light nuclei, heavy nuclei, and global nuclei.
We also refitted the SLy4 force with the group of global nuclei. The details of the optimizations and lists of parameterizations are given in Ref.\ucite{27}

%\begin{equation}
%    \left.
%    \begin{aligned}
%    E_{sym}(\rho)= & \frac{\hbar^2}{6m}\left(\frac{3\pi^2}{2}\right)^{2/3}\rho^{2/3}-\frac{1}{8}t_0(2x_0+1)\rho \\
%    & -\frac{1}{24}[3t_1x_1-t_2(4+5x_2)]\left(3\pi^2\right)^{2/3}\rho^{5/3} \\
%    & -\frac{1}{48}\sum\limits_{i}t_{3i}(2x_{3i}+1)\rho^{\sigma_i+1}
%    \end{aligned}
%    \right.
%   \label{eq}
%\end{equation}

\emph{Equilibrium of $npe\mu$ matter.}---
 Nuclear matter in neutron star is expected to be a mixture of baryons (protons and neutrons) and leptons in equilibrium with respect to weak interactions~\ucite{13,23}.
The relative fractions of $npe\mu$ components are crucial for the rate of
neutrino cooling of neutron stars. The total energy density is a function of particle densities $\rho_n, \rho_p, \rho_e, \rho_{\mu}$, and is written as:
\begin{equation}
\begin{aligned}
    \epsilon(r) &= \epsilon_b(\rho_n, \rho_p)+\rho_n m_nc^2+\rho_p m_pc^2+\epsilon_e(\rho_e)\\
    & +\epsilon_{\mu}(\rho_{\mu})+\rho_em_ec^2+\rho_{\mu}m_{\mu}c^2
    \label{eq}
     \end{aligned}
\end{equation}
Note that the charge neutrality is kept as $\rho_p=\rho_e+\rho_{\mu}$.
The equilibrium of the chemical potentials (defined as $\mu_{j}=\dfrac{\partial \epsilon}{\partial \rho_j}$) is written as:
\begin{equation}
\begin{aligned}
    \mu_n=\mu_p+\mu_e,\quad\mu_{\mu}=\mu_e
    \label{eq}
     \end{aligned}
\end{equation}
Based on the equilibrium equation, we can calculate the densities of various components at a given baryon density and thus the equation of state at  $npe\mu$ equilibrium is obtained.

\emph{Tolman-Oppenheimer-Volkov equation.}---
The TOV equation describes the static density distributions of spherical neutron star at gravitational equilibrium, which is derived from general relativity.\ucite{35,36}
The TOV differential equation in terms of pressure $P(r)$, energy density $\epsilon(r)$, and mass $M(r)$ is written as:
\begin{equation}
\begin{aligned}
    \frac{\mathrm{d}P}{\mathrm{d}r}=-\frac{GM\epsilon}{r^2c^2} (1+ \frac{P}{\epsilon})(1+ \frac{4\pi r^3 P}{ Mc^2})(1- \frac{2GM}{rc^2})^{-1} \\
    \label{eq}
     \end{aligned}
\end{equation}
The mass $M(r)$ is the total mass contained inside radius $r$, while $M(0)$=0.

With calculated EOS $P(\rho)$, the TOV equation can be numerically solved from the center with an initial density to the surface where the pressure becomes zero.
We obtain the density distribution and total mass, and the associated radius with certain boundary conditions.
It can be seen that EoS is crucial to determine the mass-radius relation of neutron stars.

\emph{Tidal deformability.}---The dimensionless tidal deformability $\Lambda$ is a functional of dimensionless compactness $\mathcal{C}=GM/Rc^2$.\ucite{9} $\Lambda$ becomes an important additional constraints on theoretical EOS, since it was first inferred from gravitational waves of GW170817.\ucite{6} The tidal deformability $\Lambda$ is obtained by solving equations as follows.\ucite{37,38}
First, a differential equation is solved to obtain $y(r)$ . $y_R$ is the value of $y(R)$ when $R$ is the NS radius.
\begin{equation}
    \frac{\mathrm{d}y(r)}{\mathrm{d}r}=-\frac{1}{r}[{y(r)}^2+y(r)F(r)+r^2Q(r)]
    \label{eqyr}
\end{equation}
The above functions $F(r)$ and $Q(r)$, related to the  total mass $M(r)$  and pressure $P(r)$, energy density $\epsilon(r)$ at radius $r$, is written as, \ucite{9}
\begin{equation}
    F(r)=\left\{1-4\pi r^2G[\epsilon-P]/c^4\right\}\left(1-\dfrac{2GM}{rc^2}\right)^{-1}
\end{equation}
\begin{equation}
    \left.
	\begin{aligned}
        Q(r)= &\left\{\frac{4\pi G}{c^4}\left[5\epsilon+9P+\frac{\epsilon+P}{\partial P/\partial \epsilon}\right]-\frac{6}{r^2}\right\}\\
        & \left(1-\frac{2GM}{rc^2}\right)^{-1} -\frac{4G^2M^2}{r^4c^4} \\
		&\left(1+\frac{4\pi r^3P}{Mc^2}\right)^2\left(1-\frac{2GM}{rc^2}\right)^{-2}
	\end{aligned}
    \right.
\end{equation}
After obtaining $y_R$, the tidal deformability can be calculated as:\ucite{37,38}
\begin{equation}
    \left.
    \begin{aligned}
        \Lambda= & \frac{16}{15}(1-2\mathcal{C})^2[2+2\mathcal{C}(y_R-1)-y_R] \\
        &\{2\mathcal{C}[6-3y_R+3\mathcal{C}(5y_R-8)] \\
        & + 4\mathcal{C}^3[13-11y_R+\mathcal{C}(3y_R-2)+2\mathcal{C}^2(1+y_R)] \\
        & +3(1-2\mathcal{C})^2[2-y_R+2\mathcal{C}(y_R-1)]\ln{(1-2\mathcal{C})}\}^{-1}
    \end{aligned}
    \right.
\end{equation}
%p{1.3cm}<{\centering}p{1.3cm}<{\centering}p{1.3cm}<{\centering}p{1.3cm}<{\centering}p{1.3cm}p<{\centering}p{1.3cm}<{\centering}

\vskip 2mm
\centerline{\footnotesize \begin{tabular}[tb]{p{7.5cm}} TABLE\ \uppercase\expandafter{\romannumeral1}.
Global calculations of binding energies of 603 even-even nuclei and charge radii of 339 nuclei are compared with experimental data.\ucite{39,40}
Calculations are performed with the original SLy4, the refitted SLy$4{'}$, and three extended SLy4E parameter sets.
The standard root-mean-square (rms) deviations of binding energies $\sigma_E$  (in MeV) and charge radii $\sigma_r$  (in fm) are listed.
\end{tabular}}
\vskip 0.5mm
\centerline{
\begin{tabular}[tbh]{lccccc}
\hline \hline
 Parameter & SLy4   &SLy$4{'}$ & global & light  & heavy  \\ \hline
 $\sigma_E(Z\leq28)$ & 2.22   &  2.832    & 2.302  & 1.426  & 2.281  \\
 $\sigma_E(Z\leq82)$ & 3.18   & 2.582     & 2.002  & 2.381  & 2.373  \\
 $\sigma_E(all)$ & 4.372  & 2.860     & 2.303  & 3.013  & 2.576  \\ \hline
% $\sigma_r(Z\leq28)$ & 0.0484 &      & 0.0436 & 0.0389 & 0.0434 \\
% $\sigma_r(Z\leq82)$ & 0.0272 &      & 0.0232 & 0.0202 & 0.0232 \\
 $\sigma_r(all)$ & 0.028  &  0.0287    & 0.0244 & 0.0203 & 0.0244 \\ \hline \hline
\end{tabular}}

\emph{Results.}---
We firstly performed global  Skyrme-Hartree-Fock+BCS calculations of 603 even-even nuclei. The obtained binding energies of 603 nuclei\ucite{39} and charge radii of 339 nuclei\ucite{40}
are compared with experimental data, shown in Table I. For SLy4 force, the overall rms of deviations in binding energies is about 4.38 MeV, which is very large.
In particular, SLy4 seriously underestimates the binding energies of heavy and superheavy nuclei remarkably.\ucite{27} This undermine its descriptions of even larger systems such as neutron stars.
We also see that overall rms of the refitted SLy4${'}$ is about 2.86 MeV, that is improved significantly.
We also performed calculations with extended SLy4 forces with an additional high-order density dependent term.
The SLy4E$_{\rm light}$ is excellent for light nuclei but its precision reduces towards heavy nuclei.  SLy4E$_{\rm heavy}$ is acceptable for both light and heavy nuclei.
SLy4E$_{\rm global}$ is further improved compared to SLy4E$_{\rm heavy}$ and can be used for global calculations of finite nuclei with a rms of 2.57 MeV.
The charge radii are generally also improved by extended SLy4 forces. The SLy4E$_{\rm light}$ was adopted for studies of collective excitations of weakly bound nuclei. \ucite{41}
 The extended Skyrme forces based on SkM$^{*}$ and UNEDF0 are very successful for
global descriptions of finite nuclei and nuclear driplines,\ucite{42,43}  but are not successful for descriptions of neutron stars.

%\centerline{\includegraphics[width=0.45\textwidth]{pressure-symmetry}
%            \label{fig:1}}
\centerline{\includegraphics[width=0.45\textwidth]{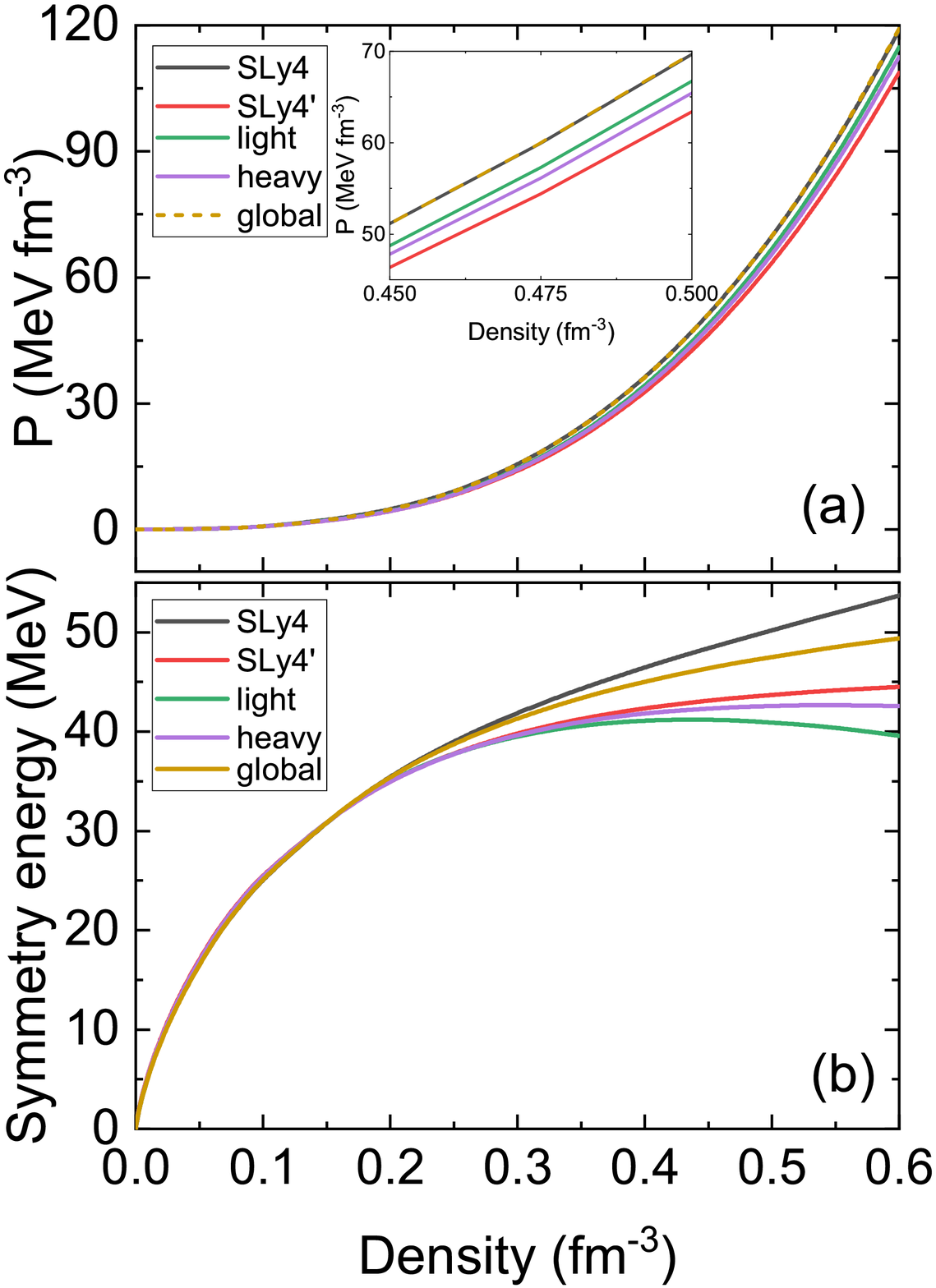}
            \label{fig:1}}
\vskip 2mm

\centerline{\footnotesize \begin{tabular}{p{7.5cm}}\bf Fig.\,1. \rm
(a) Pressure of pure neutron matter and (b) symmetry energy  as a function of nuclear density, corresponding to different Skyrme parameter sets, see text for details.
\end{tabular}}

Next we study the influences of high-order density dependent term on nuclear equation of state.
In Fig.1(a), the pressures of neutron matter by different extended SLy4 forces are shown.
For various modified SLy4 forces at the saturation point ($\rho_0$=0.16 fm$^{-3}$),  the pressure ranges from 2.2 to 2.5 MeV.
For densities lower than the saturation density, the pressures are almost the same.
In the high density region, we see SLy4$'$ gives the lowest pressure, while SLy4 and SLy4E$_{\rm global}$ gives the highest pressure.
The pressure is also related to the sound of speed in dense nuclear matter.\ucite{23}
We see the high-order density dependent term can increase the pressure at high densities. This is understandable that the repulsive
high-order term increases the pressure at high densities, which is particular important for descriptions of neutron stars.
The symmetry energies is related to EoS and  has been extensively studied by experiments of heavy ion collisions.\ucite{20}
The symmetry energy at the saturation density is around 32 MeV for SLy4 forces, while it has large uncertainties at high densities.
In our cases, it can be seen that SLy4$'$, SLy4E$_{\rm light}$, SLy4E$_{\rm heavy}$ give very soft symmetry energies at high densities.
The symmetry energies of SLy4E$_{\rm global}$ are slightly smaller than that of SLy4 results.

\vskip 0.5\baselineskip

\vskip 4mm

\centerline{\includegraphics[width=0.45\textwidth]{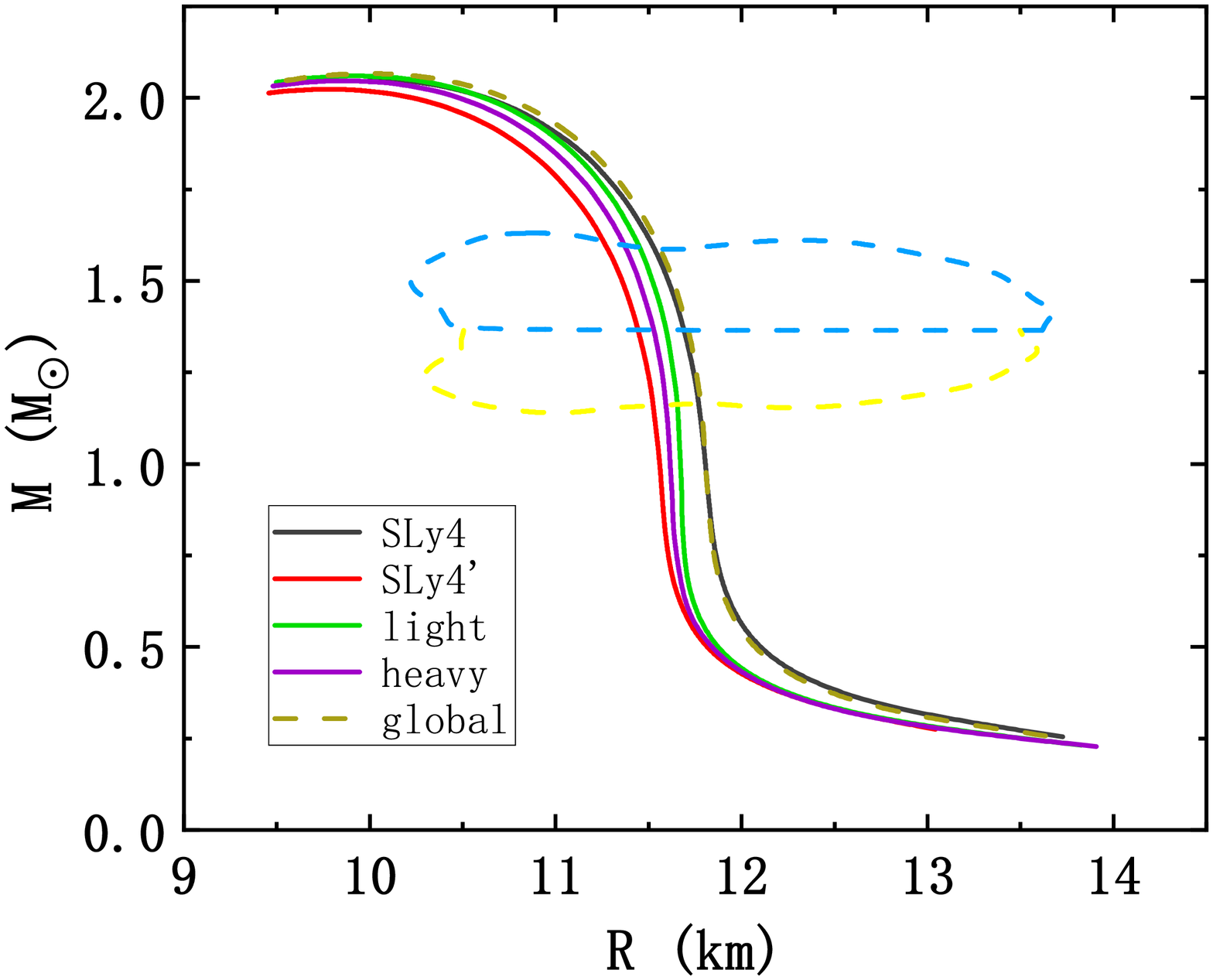}
            \label{fig:2}}

\vskip 2mm

\centerline{\footnotesize \begin{tabular}{p{7.5cm}}\bf Fig.\,2. \rm
Mass-radius relation obtained from EOS of different Skyrme parameter sets. The dashed regions correspond to constrained mass and radius of the two NSs in GW170817
at $90\%$ confidence level.\ucite{6}
\end{tabular}}

\vskip 0.5\baselineskip

We now solve TOV equation to study the properties of neutron stars with various Skyrme forces.
Note that we consider the neutron star matter is at the $npe\mu$ equilibrium, which leads to a slightly soft EoS compared to the pure neutron matter.
Fig.2 shows the calculated mass-radius relation. The crucial criteria for testing effective forces is to reproduce the observed maximum mass of neutron stars,
which is about 2 solar mass.\ucite{24} The soft EOS can not support such a large mass.
It can be seen that the maximum masses of all forces are larger than 2 solar mass.
The secret of series of successful SLy4 forces is that its $x_2$ parameter is $-1.0$.\ucite{23}
SLy4$'$ results in a slightly smaller maximum mass than that of SLy4.
The $M-R$ relation from SLy4 and SLy4E$_{\rm global}$ are almost the same.
We can see that SLy4$'$ results in the smallest radii of 11.42 km for a NS with 1.4$M_{\bigodot}$.
On the other hand, SLy4 and SLy4E$_{\rm global}$ correspond to the largest radii of 11.67 and 11.70 km, respectively.
It can be understood that the radii of 1.4 $M_{\bigodot}$ and the maximum NS mass are exactly correlated with the pressure at high densities, rather than the symmetry energy.
In contrast to precise measurements of NS masses, there are large uncertainties in measurements of NS radii.
The Bayesian analysis of GW170817 binary NSs results in a radius of 11.9$\pm$1.4 km for two NSs.\ucite{6} We see that our results agree well with the $M-R$ relation of GW170817 binary neutron stars.
Another Bayesian analysis of GW170817 provides a more stringent constraint of radii of 12.36$^{+0.52}_{-0.38}$ km and 12.32$^{+0.66}_{-0.43}$ km, respectively.\ucite{7}
%The new analysis suggests a large radii $R_{1.4}$=.
The GW170817 measurement excluded very stiff EOS.

\vskip 4mm

\centerline{\includegraphics[width=0.45\textwidth]{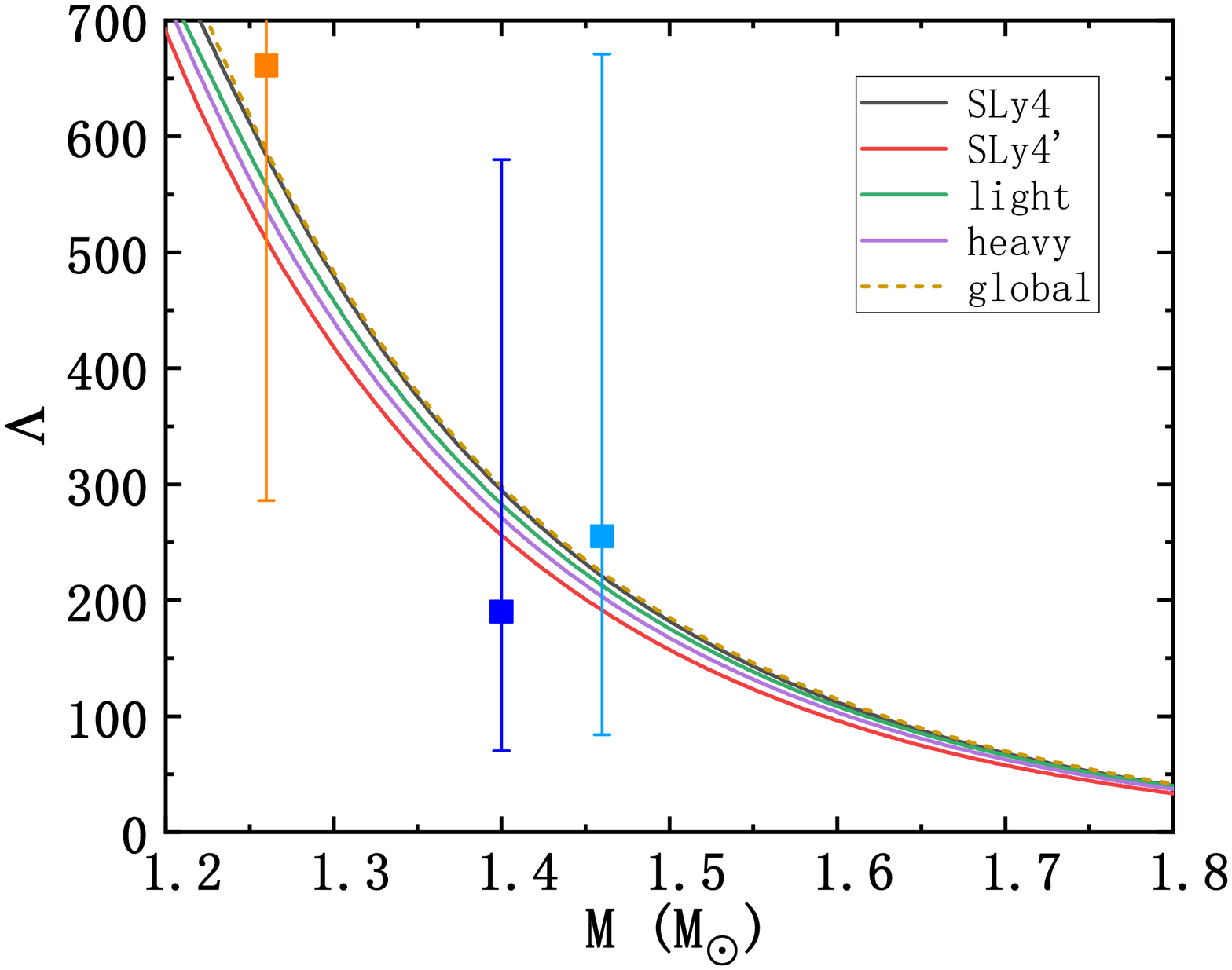}
            \label{fig:3}}

\vskip 2mm

\centerline{\footnotesize \begin{tabular}{p{7.5cm}}\bf Fig.\,3. \rm
Calculated tidal deformability $\Lambda$ as a function of NS's mass, with different Skyrme parameter sets. The inferred tidal deformability with uncertainties from Bayesian
analysis of GW180817 binary NSs are also shown. The inferred tidal deformability of 1.4 $M_{\bigodot}$ is also shown.
\end{tabular}}

\vskip 0.5\baselineskip

Finally, the calculated  tidal deformability $\Lambda$ with the five SLy4 parameter sets are shown in Fig.\,3 as a function of NS's mass.
We see that generally the tidal deformability is very large for NS with small masses.
SLy4 and SLy4E$_{\rm global}$ result in the largest tidal deformability, which is related to the largest $R_{1.4}$.
On the other hand, SLy4$'$ results in smallest tidal deformability, corresponding to the smallest $R_{1.4}$.
The differences in tidal deformability from various Skyrme forces are correlated with the $M-R$ relation.
The Bayesian analysis of the GW170817 event suggests that at $90\%$ confidence level, the tidal deformabilities $\Lambda_{1,2}$ of the binary stars are $\Lambda_1=255^{+416}_{-171}$ and $\Lambda_2=661^{+858}_{-375}$, respectively.\ucite{7}
The tidal deformability of 1.4 $M_{\bigodot}$ is estimated to be $\Lambda_{1.4}=190^{+390}_{-120}$ at $90\%$ level.\ucite{6,7}
We see that SLy4 and SLy4E$_{\rm global}$ results agree better with the observed tidal deformability of GW170817.

\emph{Conclusion.}---
The observations of GW170817 binary neutron stars, including the tidal deformability from gravitational waves for the first time,  provided a great opportunity for constraining the equation of state of dense nuclear matter.
It is known that the EoS of dense nuclear matter has large uncertainties which is related to large uncertainties of nuclear interactions at high densities.\ucite{10,44-46}
There are also extensive studies to constrain EoS or symmetry energies at high densities based on correlations with other nuclear observables.\ucite{10,44-46}
It is crucial to develop a unified and consistent description of finite nuclei and neutron stars.
We developed a series of extended SLy4 forces with an additional high-order density dependent term by fitting the properties of finite nuclei, as inspired by the Lee-Huang-Yang formula.
The original SLy4 force is widely used in studying neutron stars with the maximum mass above 2.0 solar mass but is not precise for finite nuclei.
The refitted SLy4$'$ without high-order density dependence can improve the descriptions of finite nuclei but its EOS becomes soft.
The extended SLy4E$_{\rm global}$ are satisfactory for finite nuclei and neutron stars simultaneously. Our results agree well with the GW170817 observations.
We demonstrated that the essential role of the high-order density dependent term in effective nuclear interactions at ultrahigh densities.
Our unified SLy4E$_{\rm global}$  descriptions provide a predictive model for studying neutron stars, as well as new understandings of effective interactions.
The realistic studies of neutron stars including superfluidity and non-uniform crusts are of interests in the future.
We expect that the more precise observations of gravitational waves in the future can
provide more stringent constraints on nuclear interactions in ultradense nuclear matter.

\bibliography{ref1}{}
\bibliographystyle{plain}

\section*{\Large\bf References}

\vspace*{-0.8\baselineskip}\frenchspacing

\hskip 7pt {\footnotesize

\REF{[1]} Baade W and Zwicky F 1934 {\it National Academy of Sciences} {\bf 20} 259  %DOI:10.1073/pnas.20.5.259

\REF{[2]} James M L and Madappa P 2016 {\it Physics Reports} {\bf 621} 127 %DOI:10.1016/j.physrep.2015.12.005

\REF{[3]} Oertel M, Hempel M, Kl\"ahn T and Typel S 2017 {\it Rev. Mod. Phys.} {\bf 89} 015007 %DOI:10.1103/RevModPhys.89.015007

\REF{[4]} Rikovska S J, Miller J C, Koncewicz R, Stevenson P D and Strayer M R 2003 {\it Phys. Rev. C.} {\bf 68} 034324 %DOI:10.1103/PhysRevC.68.034324

\REF{[5]} Abbott B P $et\ al.$ 2017 {\it Phys. Rev. Lett.} {\bf 119} 161101 %DOI:10.1103/PhysRevLett.119.161101

\REF{[6]} Abbott B P $et\ al.$ 2018 {\it Phys. Rev. Lett.} {\bf 121} 161101 %DOI:10.1103/PhysRevLett.121.161101

\REF{[7]} Fasano M, Abdelsalhin T, Maselli A and Ferrari V 2019 {\it Phys. Rev. Lett.} {\bf 123} 141101 %DOI:10.1103/PhysRevLett.123.141101

\REF{[8]} Abbott B P $et\ al.$ 2019 {\it Phys. Rev. X} {\bf 9} 011001 %DOI:10.1103/PhysRevX.9.011001

\REF{[9]} Hinderer T 2008 {\it Astrophys. J.} {\bf 677} 1216 %DOI:10.1086/533487

\REF{[10]} Burgio G F and Vida\~na I 2018 {\it Universe} {\bf 6} 119 %DOI:10.3390/universe6080119

\REF{[11]} Douchin F and Haensel P 2001 {\it Astro. Astrophys.} {\bf 380} 151 %DOI:10.1051/0004-6361:20011402

\REF{[12]} James M L and Madappa P 2001 {\it Astrophys. J.} {\bf 550} 426 %DOI:10.1086/319702

\REF{[13]} Stone J R, Stevenson P D, Miller J C and Strayer M R 2002 {\it Phys. Rev. C.} {\bf 65} 064312 %DOI:10.1103/PhysRevC.65.064312

\REF{[14]} Ishii N, Aoki S and Hatsuda T 2007 {\it Phys. Rev. Lett.} {\bf 99} 022001 %DOI:10.1103/PhysRevLett.99.022001

\REF{[15]} Tews I, Davoudi Z, Ekstrom A, Holt J D and Lynn J E 2020 {\it J. Phys. G.} {\bf 47} 103001 %DOI:10.1088/1361-6471/ab9079

\REF{[16]} Dutra M, Louren\ifmmode \mbox{\c{c}}\else \c{c}\fi{}o O, S\'aMartins J S, Delfino, A, Stone J  and Stevenson P D 2012 {\it Phys. Rev. C.} {\bf 85} 035201 %DOI:10.1103/PhysRevC.85.035201

\REF{[17]} Li A, Zhu Z Y, Zhou E P, Dong J M, Hu J N and Xia C J 2020 {\it JHEAp} {\bf 28} 19 %DOI:10.1016/j.jheap.2020.07.001

\REF{[18]} Tamii, A. $et\ al.$ 2011 {\it Phys. Rev. Lett.} {\bf 107} 062502 %DOI:10.1103/PhysRevLett.107.062502

\REF{[19]} Abrahamyan $et\ al.$ 2012 {\it Phys. Rev. Lett.} {\bf 108} 112502 %DOI:10.1103/PhysRevLett.108.112502

\REF{[20]} Tsang M B, Zhang Y X, Danielewicz P, Famiano M, Li Z X, Lynch W G and Steiner A W 2009 {\it Phys. Rev. Lett.} {\bf 102} 122701; %DOI:10.1103/PhysRevLett.102.122701
 Danielewicz P, Lacey R and Lynch W G 2002 {\it Science} {\bf 298} 1592; %DOI:10.1126/science.1078070
 Li B A, Chen L W and Che M K 2008 {\it Phys. Rep.} {\bf 464} 113; %DOI:10.1016/j.physrep.2008.04.005
Xiao Z G, Li B A, Chen L W, Yong G C, and Zhang M 2009 {\it Phys. Rev. Lett.} {\bf 102} 062502
%Chen L W 2015 {\it EPJ Web Conf.} {\bf 88} 00017 %DOI:10.1051/epjconf/20158800017

\REF{[21]} Skyrme T H R 1956 {\it Philos. Mag} {\bf 1} 1043 %DOI:10.1080/14786435608238186

\REF{[22]} Vautherin D and Brink D M 1972 {\it Phys. Rev. C.} {\bf 5} 626 %DOI:10.1103/PhysRevC.5.626

\REF{[23]} Chabanat E, Bonche P, Haensel P, Meyer J and Schaeffer R 1997 {\it Nuclear Physics A} {\bf 627} 710 %DOI:10.1016/S0375-9474(97)00596-4

\REF{[24]} Chabanat E, Bonche P, Haensel P, Meyer J and Schaeffer R 1998 {\it Nuclear Physics A} {\bf 635} 231 %DOI:10.1016/S0375-9474(98)00180-8

\REF{[25]} Antoniadis John $et\ al.$ 2013 {\it Science} {\bf 340} 1233232 %DOI:10.1126/science.1233232

\REF{[26]} Erler J, Horowitz C J, Nazarewicz W, Rafalski M and Reinhard P G 2013 {\it Phys. Rev. C.} {\bf 87} 044320 %10.1103/PhysRevC.87.044320

\REF{[27]} Xiong X Y, Pei J C and Chen W J 2016 {\it Phys. Rev. C.} {\bf 93} 024311 %DOI:10.1103/PhysRevC.93.024311

\REF{[28]} Negele J W 1982 {\it Rev. Mod. Phys.} {\bf 54}  913 %DOI:10.1103/RevModPhys.54.913

\REF{[29]} Grange P, Lejeune A, Martzolff M and Mathiot J F 1989 {\it Phys. Rev. C.} {\bf 40} 1040 %DOI:10.1103/PhysRevC.40.1040

\REF{[30]} Huang K and Yang C N 1957 {\it Phys. Rev.} {\bf 105} 767 %DOI:10.1103/PhysRev.105.767

\REF{[31]} Lee T D and Yang C N 1957 {\it Phys. Rev.} {\bf 105}  1119 %DOI:10.1103/PhysRev.105.1119

\REF{[32]} DeDominicis C and Martin P C 1957 {\it Phys. Rev.} {\bf 105} 1419 %DOI:10.1103/PhysRev.105.1419

\REF{[33]} Altmeyer A, Riedl S, Kohstall C, Wright M J, Geursen R, Bartenstein M, Chin C, Denschlag J H and Grimm R 2017 {\it Phys. Rev. Lett.} {\bf 98} 040401 %DOI:10.1103/PhysRevLett.98.040401

\REF{[34]} Agrawal B K, Dhiman S K and Kumar R 2006 {\it Phys. Rev. C.} {\bf 73} 034319 %DOI:10.1103/PhysRevC.73.034319

\REF{[35]} Tolman R C 1939 {\it Phys. Rev.} {\bf 55} 364 %DOI:10.1103/PhysRev.55.364

\REF{[36]} Oppenheimer J R and Volkoff G M 1939 {\it Phys. Rev.} {\bf 55} 374 %DOI:10.1103/PhysRev.55.374

\REF{[37]} Hu J N, Bao S S, Zhang Y, Nakazato K I, Sumiyoshi K and Shen H 2020 {\it Prog. Theo.  Expt. Phys.} {\bf 2020} 10.1093/ptep/ptaa016 %DOI:10.1093/ptep/ptaa016

\REF{[38]} Hinderer T, Lackey B D, Lang R N and Read J S 2010 {\it Phys. Rev. D.} {\bf 81} 123016 %DOI:10.1103/PhysRevD.81.123016

\REF{[39]} Wang M, Audi G, Kondev G G, Huang W J, Naimi S and Xu X 2017 {\it Chin. Phys. C.} {\bf 41} 030003 %DOI:10.1088/1674-1137/41/3/030003

\REF{[40]} Angeli I and Marinova K P 2013 {\it Atomic Data and Nuclear Data Tables} {\bf 99} 69 %DOI:10.1016/j.adt.2011.12.006

\REF{[41]} Wang K, Kortelainen M and Pei J C 2017 {\it Phys. Rev. C.} {\bf 96} 031301  %DOI:10.1103/PhysRevC.96.031301

\REF{[42]} Zuo Z W, Pei J C, Xiong X Y and Zhu Y 2018 {\it Chin. Phys. C.} {\bf 42} 064106  %DOI:10.1088/1674-1137/42/6/064106

\REF{[43]} Chai Q Z, Pei J C, Fei N and Guan D W 2020 {\it Phys. Rev. C.} {\bf 102} 014312  %DOI:10.1103/PhysRevC.102.014312

\REF{[44]} Malik T, Agrawal B K, De J N, Samaddar S K, Provid\^encia C, Mondal C and Jha T K 2019 {\it Phys. Rev. C.} {\bf 99} 052801  %DOI:10.1103/PhysRevC.99.052801

\REF{[45]} Tsang C Y, Tsang M B, Danielewicz P, Fattoyev F J, Lynch W G 2019 {\it Phys. Lett. B.} {\bf 796} 1  %DOI:10.1016/j.physletb.2019.05.055

\REF{[46]} Zhang Y X, Liu M, Xia C J, Li Z X and Biswal S K 2020 {\it Phys. Rev. C.} {\bf 101} 034303  %DOI:10.1103/PhysRevC.101.034303

%\REF{[23]} Dutra, M. and Louren\ifmmode \mbox{\c{c}}\else \c{c}\fi{}o, O. and S\'a Martins, J. S. and Delfino, A. and Stone, J. R. and Stevenson, P. D. 2019 {\it Phys. Rev. C.} {\bf 85} 035201 %DOI:10.1103/PhysRevC.85.035201

%\REF{[24]}  Available at http://jblevins.org/mirror/amiller/simann.f90

%\REF{[27]} S. E. Thorsett and Deepto Chakrabarty 1999 {\it Astrophys. J.} {\bf 512} 288 %DOI:10.1086/306742

}

\end{multicols}
\end{document}